\documentclass[%
reprint, superscriptaddress,
nofootinbib, preprintnumbers,
 amsmath,amssymb,
 aps,
pra,
]{revtex4}
\usepackage{amsmath}
\usepackage{amsthm}
\usepackage{physics}
\usepackage{graphicx}
\usepackage{bm}
\usepackage{color} 
\usepackage{CJK}

\def\la{{\langle}}
\def\ra{{\rangle}}

\newcommand{\beq}{\begin{equation}}
\newcommand{\eeq}{\end{equation}}
\newcommand{\beqa}{\begin{eqnarray}}
\newcommand{\eeqa}{\end{eqnarray}}

\begin{document}
\title{Symmetries and invariants for non-Hermitian Hamiltonians}
\author{M. A. Sim\'on Mart\'\i nez$^{1}$, A. Buend\'\i a$^{1}$,  and J. G. Muga}
\affiliation{%
Departamento de Qu\'{\i}mica F\'{\i}sica, UPV/EHU, Apdo.
644, 48080 Bilbao, Spain}

\begin{abstract}
We discuss Hamiltonian symmetries and invariants for quantum systems driven by
non-Hermitian Hamiltonians. For time-independent Hermitian Hamiltonians, a unitary or antiunitary transformation $AHA^\dagger$ that leaves the Hamiltonian $H$
unchanged represents a symmetry of the Hamiltonian, which implies the commutativity $[H,A]=0$, and a conservation law,
namely the invariance of expectation values of $A$.
For non-Hermitian Hamiltonians, $H^\dagger$ comes into play as a distinct operator that complements $H$ in generalized unitarity relations.
The above description of symmetries has to be extended to include also $A$-pseudohermiticity
relations of the form
$AH=H^\dagger A$. A superoperator formulation of Hamiltonian symmetries is provided and exemplified for Hamiltonians
of a particle moving in one-dimension considering the set of $A$ operators
forming Klein's 4-group: parity, time-reversal, parity\&time-reversal, and unity.  The link between symmetry and conservation laws is
discussed and shown to be more subtle for non-Hermitian than for Hermitian Hamiltonians.
\end{abstract}

\maketitle

\section{Introduction}
The intimate link between invariance and symmetry is well studied and understood
for
Hermitian Hamiltonians but non-Hermitian Hamiltonians pose some interesting
conceptual and formal challenges.
Non-Hermitian Hamiltonians arise naturally in quantum systems as effective interactions for a subsystem. These Hamiltonians may be proposed phenomenologically or may be
found exactly or approximately by applying Feshbach's projection technique to describe the dynamics in the subsystem \cite{Feschbach(1958),Ruschhaupt(2004)}.
It is thus  important to understand
how common concepts for Hermitian Hamiltonians such as ``symmetry'',
``invariants'', or ``conservation laws'' generalize.
A lightning review of concepts and formal relations for a time-independent Hermitian Hamiltonian $H$ will be helpful as the starting point to
address generalizations for a non-Hermitian $H$. Unless stated otherwise, $H$ is time-independent in the following.
In quantum mechanics $A$ (unitary or antiunitary) represents a symmetry of the Hamiltonian if
\begin{equation}
A^\dagger HA=H,
\label{symme}
\eeq
so that
\beq
[H,A]=0,
\label{conmu}
\eeq
and thus $A$ (which we assume to be time-independent) represents also a conserved quantity,
\beq
\la \psi(t),A\psi(t)\ra=\la \psi(0),A \psi(0)\ra,
\label{conser}
\eeq
where $|\psi(t)\ra=U(t)|\psi(0)\ra$ is the time-dependent wave function
satisfying the  Schr\"odinger equation
\begin{equation}
    i\hbar\partial_t\ket{\psi(t)} = H \ket{\psi(t)},
    \label{eq:ScrdngrEqn}
\end{equation}
and $U(t)=e^{-iHt/\hbar}$ is the unitary evolution operator from $0$ to $t$,
$U(t)U^\dagger(t)=U^\dagger(t)U(t)=1$. Backwards evolution in time from $t$ to 0 is represented by $U(-t)=U(t)^\dagger$ so that the initial state is recovered by
a forward and backward sequence,
$U(t)^{\dagger}U(t)|\psi(0)\ra=|\psi(0)\ra$.

Equation (\ref{conser}) is mostly significant for a linear $A$. If $A$ is antilinear
only the modulus is relevant,
as the result changes if we multiply the state by a unit modulus phase factor, $\psi(0)\to e^{i\phi}\psi(0)$. This ambiguity
does not mean at all that antilinear symmetries do not have physical consequences.
They affect, for example, selection rules for possible transitions.

More generally, time-independent operators $A$ satisfying (\ref{conmu}), fullfill (\ref{conser}) without the need to be unitary or antiunitary,
and represent also invariant quantities.
A further property from (\ref{conmu}) is that if $|\phi_E\ra$ is an eigenstate of $H$ with (real) eigenvalue $E$, then
$A|\phi_E\ra$ is also an eigenstate of $H$ with the same eigenvalue.
%
%
%
\section{Dual character of $H$ and $H^\dagger$}
Defining $\widehat{U}(t)=e^{-iH^\dagger t/\hbar}$, we find the generalized unitarity relations
$U(t)\widehat{U}^\dagger(t)=\widehat{U}^\dagger U(t)=1$. Backwards evolution with $H^\dagger$
compensates the changes induced forwards by $H$. Similar generalized unitarity relations exist for the scattering
$S$ matrix (for evolution with $H$) and the corresponding $\widehat{S}$ (for evolution with $H^\dagger$), with
important physical consequences discussed e.g. in \cite{Muga(2004),Ruschhaupt(2017)}.
%

%
%
%
%
Now consider the following two formal generalizations of the element $\la \psi(t),A\psi(t)\ra$ in Equation (\ref{conser}),
\beqa
&&\la e^{-iH^\dagger t/\hbar} \psi(0), A e^{-iHt/\hbar} \psi(0)\ra=\la \widehat{\psi}(t),A\psi(t)\ra,
\label{sec}
\\
&&\la e^{-iH t/\hbar}  \psi(0), A e^{-iHt/\hbar} \psi(0)\ra=\la \psi(t), A \psi(t)\ra,
\label{pri}
\eeqa
and the  generalizations of (\ref{conmu})
\beqa
AH&=&HA,
\label{usu}
\\
AH&=&H^\dagger A.
\label{unu}
\eeqa
We name (\ref{unu}) $A$-pseudohermiticity of $H$ \cite{Mostafazadeh(2010)}.
(This is here a formal definition that does not presupose any  further property
on $A$.)
Up  to normalization, which will be discussed in the following section, Equation (\ref{pri}) corresponds to the usual rule to
define expectation values, whereas (\ref{sec}), where $|\widehat{\psi}(t)\ra\equiv e^{-iH^\dagger t/\hbar} |\psi(0)\ra$,
is unusual, and its physical meaning is not obvious. Note however that, for linear $A$,
$AH=HA$ implies the conservation of the unusual quantity (\ref{sec}), whereas
$A$-pseudohermiticity $AH=H^\dagger A$ implies the conservation of the usual quantity (\ref{pri}) \cite{Vitanov(2016)}.
At this point we might be tempted to discard (\ref{usu}) as less useful or significant physically.
This is however premature for several reasons. One is the following (others will be seen in Sections 4 to 6):
Unlike Hermitian Hamiltonians, non-Hermitian ones may have generally different
right and left eigenvectors. We assume the existence of the resolution
\beq
H=\sum_j |\phi_j\ra E_j \la\widehat{\phi}_j|,
\label{res}
\eeq
where the $E_j$ may be complex and where
\beqa
H|\phi_j\ra=E_j|\phi_j\ra, \;\;\;
H^\dagger|\widehat{\phi}_j\ra=E_j^* |\widehat{\phi}_j\ra.
\eeqa
We have used a simplifying notation assuming a discrete spectrum, but a continuum
part could be treated similarly by adding integrals and continuum-normalized states.
Note that left eigenstates of $H$ are right eigenstates of $H^\dagger$
with a complex conjugate eigenvalue. If $|\phi_j\ra$ is  a right eigenstate of $H$
with eigenvalue $E_j$, Equation (\ref{usu}) implies that
$A|\phi_j\ra$ is also a right  eigenstate of $H$, with the
same eigenvalue if $A$ is linear, and with the complex conjugate eigenvalue $E_j^*$ if $A$ is antilinear.
Instead, Equation (\ref{unu}) implies that $A|\phi_j\ra$ is a right eigenstate of $H^\dagger$
with eigenvalue $E_j$ for $A$ linear or $E_j^*$ for $A$ antilinear, or a left eigenstate of $H$ with eigenvalue $E_j^*$ for $A$ linear, or $E_j$ for $A$ antilinear.
As right and left eigenvectors must be treated on equal footing, since both are needed for the
resolution (\ref{res}), this argument points at a similar importance of the relations (\ref{usu}) and (\ref{unu}).
\section{Time evolution for normalized states}
For a quantum system following the  Schr\"odinger Equation (\ref{eq:ScrdngrEqn})
with $H$ non-Hermitian, in general the evolution will not  be unitary and
the norm $N_{\psi}(t) \equiv \braket{\psi(t)}{\psi(t)}$ is not conserved.
We shall asume the initial condition $N_{\psi}(0)=1$. Using Equation \eqref{eq:ScrdngrEqn}, the rate of change of the norm is
\begin{equation}
    \partial_t \braket{\psi(t)}{\psi(t)} = \frac{1}{i\hbar}\bra{\psi(t)}H-H^\dagger\ket{\psi(t)} \neq 0.
    \label{eq:NormChange}
\end{equation}
{\it{Expectation Values.}}
We now restrict the discussion to linear observables $A$. Since the state of the system is not normalized to 1
for $t>0$, the expectation value formula has to take into account the norm explicitly,
\begin{equation}
  \expval{A}(t) = \frac{\ev{A}{\psi(t)}}{\braket{\psi(t)}}.
  \label{eq:ExpectValue}
\end{equation}
Using Equations (\ref{eq:ScrdngrEqn}) and (\ref{eq:NormChange}) the rate of change of the expectation value of $A$ is
\begin{equation}
  \partial_t \ev{A}(t) = \frac{1}{i\hbar}\frac{\braket{\psi(t)}\ev{AH-H^\dagger A}{\psi(t)}-\ev{H-H^\dagger}{\psi(t)}\ev{A}{\psi(t)}}{\braket{\psi(t)}^2}.
  \label{eq:ExpectValueChange}
\end{equation}
For Hermitian Hamiltonians the commutation of $A$ and $H$ leaves  the expectation values of $A$ invariant.
For non-Hermitian Hamiltonians the symmetry Equation (\ref{unu})
applied to Equation \eqref{eq:ExpectValueChange} gives
\begin{equation}
\partial_t \ev{A}(t) = \frac{-1}{i\hbar}\frac{\ev{H-H^\dagger}{\psi(t)}\ev{A}{\psi(t)}}{\braket{\psi(t)}^2}.
\label{eq:ExpectValueChangeSymmetry}
\end{equation}
If we use Equations \eqref{eq:NormChange} and \eqref{eq:ExpectValue} in Equation \eqref{eq:ExpectValueChangeSymmetry},
\beqa
\frac{\ev{A}}{\braket{\psi(t)}} \partial_t\braket{\psi(t)} &=& -  \partial_t \ev{A} ,
\\
\ev{A}\braket{\psi(t)} &=& \text{Constant}.
\eeqa
Applying the initial condition $\braket{\psi(0)}=1$,
\begin{equation}
  \ev{A}(t) = \frac{\ev{A}(0)}{\braket{\psi(t)}},
  \label{eq:ExpectValueScaling}
\end{equation}
so the expectation value of an $A$ that obeys
$AH = H^\dagger A$,  is simply rescaled by the norm of the wave function as it increases or decreases.

{\it{Lower bound on the norm of the wave function.}}
The symmetry condition $AH = H^\dagger A$ may set lower bounds to the norm along the dynamical process.
Consider a linear observable $A$ with real eigenvalues $\{a_i\}$ bounded by ${max\{\abs{a_i}\}}$.
Then, the expectation values satisfy $\abs{\ev{A}}\leq max\{\abs{a_i}\}$. If we use the result in \eqref{eq:ExpectValueScaling} we get
\begin{equation}
  \braket{\psi(t)} \geq \frac{\abs{\ev{A}(0)}}{max\{\abs{a_i}\}}.
  \label{eq:LowerBound}
\end{equation}
Equation \eqref{eq:LowerBound} bounds the norm of the state due to symmetry conditions.
A remarkable case is parity pseudohermiticity, $\Pi H = H^\dagger \Pi$, where the parity operator acts on the position eigenstates as $\Pi\ket{x} = \ket{-x}$ and has eigenvalues $\{-1,1\}$. Under this symmetry, Equation \eqref{eq:LowerBound} gives
\begin{equation}
  \braket{\psi(t)} \geq \abs{\ev{\Pi}(0)},
  \label{eq:LowerBoundParity}
\end{equation}
where $\ev{\Pi}(0)$ is the expectation value of the state at $t=0$.
\section{Generic symmetries}
We postulate that both (\ref{usu}) and (\ref{unu}), for $A$ unitary or antiunitary,  are symmetries of the Hamiltonian.
A superoperator framework helps to understand why (\ref{unu}) also represents a symmetry.
Let us define the superoperators ${\cal L}_A(\cdot) \equiv A^\dagger(\cdot) A$, ${\cal L}_\dagger (\cdot)\equiv (\cdot)^\dagger$ and ${\cal L}_{A,\dagger}(\cdot)\equiv {\cal L}_A\left({\cal L}_\dagger (\cdot)\right) = {\cal L}_\dagger\left({\cal L}_A (\cdot)\right) $. For  linear operators $B$ and a complex number $a$ they satisfy
\beqa
{\cal L}_A (a B)&=& a {\cal L}_A(B),\;\;\,  A\, {\rm unitary},
\\
{\cal L}_A (a B)&=& a^* {\cal L}_A(B),\;  A\, {\rm antiunitary},
\\
{\cal L}_{\dagger} (a B)&=& a^* {\cal L}_\dagger(B),
\\
{\cal L}_{A,\dagger} (a B)&=& a^* {\cal L}_{A,\dagger} (B),\;
A\, {\rm{unitary}},
\\
{\cal L}_{A,\dagger} (a B)&=& a {\cal L}_{A,\dagger} (B),\;
A\, {\rm{unitary}}.
\eeqa
As the product of two antilinear operators is a linear operator,
the resulting operators (on the right hand sides) are linear in all cases,
independently of the linearity or antilinearity of $A$. This should not be confused with the linearity or antilinearity
of the superoperators ${\cal L}$ that may be checked by the invariance (for a linear superoperator) or complex conjugation (for an antilinear
superoperator) of the constant $a$.
Using the scalar product for linear operators $F$ and $G$,
\beq
\la\la F,G\ra\ra\equiv Tr (F^\dagger G),
\eeq
we find the adjoints,
\beqa
{\cal L}_A^\dagger(\cdot)&=& {\cal L}_{A^\dagger}(\cdot)\equiv A(\cdot)A^\dagger,
\\
{\cal L}_{\dagger}^\dagger(\cdot) &=&{\cal L}_{\dagger}(\cdot),
\\
{\cal L}_{A,\dagger}^\dagger (\cdot) &=& {\cal L}_{A^\dagger, \dagger}(\cdot)\,,
\eeqa
where $\la\la F,{\cal L}^\dagger G\ra\ra=\la\la G,{\cal L} F\ra\ra^*$ for ${\cal L}$ linear
and $\la\la F,{\cal L}^\dagger G\ra\ra=\la\la G,{\cal L} F\ra\ra$ for ${\cal L}$ antilinear.

All the above transformations are unitary or antiunitary (in a superoperator sense),
${\cal L}^\dagger= {\cal L}^{-1}$,
and they keep ``transition probabilities'' among
two states represented by density operators
$\rho_1$ and $\rho_2$ invariant, namely
\begin{equation}
\la\la {\cal L}\rho_1,  {\cal L} \rho_2\ra\ra = \la\la \rho_1,\rho_2\ra\ra.
\label{Wigner}
\end{equation}
Due to the Hermicity of the density operators, $\la\la \rho_1,\rho_2\ra\ra$ is a real number (both for unitary or antinuitary ${\cal{L}}$). This result is reminiscent of Wigner's theorem, originally formulated for pure states \cite{Wigner(1931)},
but considering a more general set of states and transformations.

We conclude that all of the above ${\cal L}$ superoperators may represent symmetry transformations and, in particular,
Hamiltonian symmetries if they leave the Hamiltonian invariant, namely,
${\cal L} H= H$. The following section demonstrates this for the set of symmetry transformations
that leave Hamiltonians for a particle in one dimension invariant, making use of transposition, complex conjugation, and inversion of
coordinates or momenta.

As for the connection between symmetries and conservation laws, the results of the previous sections apply. It is possible
to find quantities that on calculation remain invariant, but they are not necessarily physically significant.

\section{Example of physical relevance of the relations $AH=HA$, $AH=H^\dagger A$ as symmetries}
In this section we exemplify the above general formulation of Hamiltonian symmetries
for Hamiltonians
of the form $H_0+V$ corresponding to a  particle of mass $m$ moving in one dimension, where $H_0=P^2/(2m)$
is the kinetic energy, $P$ the  momentum operator, and $V$ is a generic potential that may be
non-Hermitian and non-local (non-local means that matrix elements in coordinate representation, $\la x|V|y\ra$, may be non-vanishing for
$x\ne y$).
We assume that $H$ is diagonalizable, possibly with discrete and continuum parts.
By inspection of  Table \ref{table2}, one finds a set of possible Hamiltonian symmetries described by the eight relations of the second column. They amount to the invariance of the Hamiltonian with respect to the
transformations represented by the superoperators in the third column. In coordinate or momentum representation, see the last two columns,
each symmetry amounts to the invariance of the potential matrix elements with respect to some combination
of transposition, complex conjugation and inversion (of coordinates or momenta). ($H_0$ is invariant with respect to the eight transformations.)

The eight superoperators form the elementary abelian group of order eight \cite{Rose(2009)}, with a minimal set of three
generators
${\cal L}_\dagger, {\cal L}_\Pi, {\cal L}_\Theta$, from which all elements may be formed by multiplication (i.e., successive
application). $\Theta$ is the antilinear time-reversal operator.
Note that no other transformation is possible
that leaves the Hamiltonian invariant making only use of transposition, complex conjugation, inversion, and their combinations.
The eight superoperators may also be found by the generating set $\{{\cal L}_A\}, {\cal L}_\dagger$,
where  $A$ is one of the elements of Klein's 4-group
$\{1, \Theta,\Pi,\Pi\Theta\}$.  These four operators commute. Moreover they are Hermitian and  equal to their own inverses.
The superoperators in the third column may be classified as antiunitary (symmetries II, IV, V, and VII)
and unitary (symmetries I, III, VI, and VIII).

In \cite{Ruschhaupt(2017)} these symmetries are exploited to find selections rules that allow or disallow certain asymmetries
in the reflection or transmission amplitudes for right and left incidence, a relevant information to implement
microscopic asymmetrical devices such as
diodes or rectifiers in quantum circuits \cite{Ruschhaupt(2004b)}.

\begin{table}
\begin{tabular}{ccccc}
Code & Symmetry&  Superoperator & $\la x|V|y\ra=$&    $\la p|V|p'\ra=$
\\
\hline
I & $1H=H1$ &  ${ {\cal L}_1}$  &  $\la x|V|y\ra$ &  $\la p|V|p'\ra$
\\
II & $1H=H^\dagger 1$ &  ${\cal L}_\dagger$ & $\la y|V|x\ra^*$ &   $\la p'|V|p\ra^*$
\\
III & $\Pi H=H\Pi$ & ${\cal L}_\Pi$  & $\la -x|V|-y\ra$ &   $\la -p|V|-p'\ra$
\\
IV & $\Pi H=H^\dagger \Pi$ & ${\cal L}_{\Pi,\dagger}$  & $\la -y|V|-x\ra^*$ &   $\la -p'|V|-p\ra^*$
\\
V & $\Theta H=H\Theta$ &  ${\cal L}_\Theta$ & $\la x|V|y\ra^*$ &   $\la -p|V|-p'\ra^*$
\\
VI & $\Theta H=H^\dagger\Theta$ &  ${\cal L}_{\Theta,\dagger}$ & $\la y|V|x\ra$  & $\la -p'|V|-p\ra$
\\
VII & $\Theta\Pi H=H\Theta \Pi$ &  ${\cal L}_{\Pi,\Theta}$ & $\la -x|V|-y\ra^*$ &  $\la p|V|p'\ra^*$
\\
VIII& $\Theta\Pi H=H^\dagger \Pi\Theta$ & ${\cal L}_{\Pi,\Theta,\dagger}$ & $\la -y|V|-x\ra$ &  $\la p'|V|p\ra$
\\
\end{tabular}
\caption{Symmetries of the potential dependent on the commutativity or pseudo-hermiticity of $H=H_0+V$ with the elements of
Klein's 4-group  $\{1,\Pi,\Theta,\Pi\Theta\}$ (second column). Each symmetry has a roman number code in the first column.
Each symmetry may also be regarded as the invariance of the potential with respect to the transformations represented by superoperators
${\cal L}$ in the third column. The kinetic part $H_0$ is invariant in all cases.
In coordinate (third column) or momentum representation (fourth column), the eight transformations
correspond to all possible combinations of transposition, complex conjugation, and inversion. \label{table2}}
\end{table}

\section{Discussion}
The relations between invariance and symmetry are often emphasized,
but for non-Hermitian Hamiltonians, which occur naturally as effective interactions, they become more complex
and subtle than for Hermitian Hamiltonians.
We have discussed these relations for time-independent Hamiltonians.

For time-dependent Hamiltonians additional  elements are needed.
In 1969, Lewis and Riesenfeld \cite{LR} showed that the motion of a system subjected to time-varying forces admits a simple decomposition into elementary, independent motions characterized by constant values of some quantities (eigenvalues of the invariant).
In other words, the dynamics is best
understood, and is most economically described, in terms of invariants even for time-dependent Hamiltonians.
In fact the powerful link between
forces and invariants can be used in reverse order to inverse engineer from the invariant associated with some desired dynamics the necessary driving forces.

Time-dependent non-Hermitian Hamiltonians require a specific analysis and will be treated in more detail
elsewhere.
However we briefly advance here some important differences with the time-independent Hamiltonians.
Invariants for Hermitian time-dependent Hamiltonians obey the
invariance condition
\beq
\frac{\partial I(t)}{\partial t}-\frac{1}{i\hbar}[H(t),I(t)]=0,
\label{gene}
\eeq
so that $\frac{d}{dt}\la {\psi}(t)|I(t)|\psi(t)\ra=0$ for states $\psi(t)$ that evolve  with $H(t)$ (we assume that the invariant is linear).
In general the  operator $I(t)$ may depend on time and the invariant quantity is the expectation
value $\la \psi(t)|I(t)|\psi(t)\ra$.
In this context a  Hamiltonian symmetry, defined
by the commutativity of $A$ with $H$ as in (\ref{usu}) does not lead necessarily to a  conservation
law, unless $A$ is time independent.

Invariant operators  are useful to express the dynamics of the state $\psi(t)$ in terms of
superpositions of their  eigenvectors with constant coefficients \cite{LR}; also to do inverse engineering, as in shortcuts to adiabaticity,
so as to find $H(t)$ from the desired dynamics \cite{Ibanez(2011),Torrontegui(2013)}.

$I(t)$ may be formally defined by (\ref{gene}) for non-Hermitian Hamiltonians too, and its
roles to provide a basis for useful state decompositions and inverse engineering are still applicable \cite{Ibanez(2011)}.
Note however that in this context $I(t)$ is not invariant in an ordinary sense, but rather
\beq
\frac{d}{dt}\la \widehat{\psi}(t)|I(t)|\psi(t)\ra=0.
\eeq
The alternative option, yet to be explored for inverse engineering the Hamiltonian,
is to consider (linear) operators $I'(t)$ such that
\beq
\frac{\partial I'(t)}{\partial t}-\frac{1}{i\hbar}[H(t)^\dagger I'(t)-I'(t)H(t)]=0,
\label{gene2}
\eeq
and thus  $\frac{d}{dt}\la \psi(t)|I'(t)|\psi(t)\ra=0$.


\vspace{6pt}

{\it{Acknowledgments}---}. {We thank A. Kiely for many useful comments.
This research was funded by Basque Country Government (grant number IT472-10), and MINECO/FEDER,UE (grant number FIS2015-67161-P).}

\end{document}